\documentstyle[aps,pre,psfig,preprint]{revtex}

\tighten

\begin{document}

\title{Statistics of citation networks}

\author{Alexei Vazquez}

\address{International School for Advanced Studies, via Beirut 4, 34014
Trieste, Italy (vazquez@sissa.it)}

\maketitle

\begin{abstract} 
The out-degree distribution of citation networks is investigated. Statistical data
of the number of papers cited within a paper (out-degree) for different journals in
the period 1991-1999 is reported. The out-degree distribution is characterized by a
maximum at intermediate out-degrees. At the left of the maximum  there are strong
fluctuations from journal to journal while is quite universal at the right, with two
classes of journals.  These two classes are associated with the existence or not of
a restriction in the maximum number of pages per paper. The shape of the out-degree
distribution does not change appreciable from period to period, but the average
out-degree is observed to increase logarithmically with the number of published
papers. These features are modeled using a recursive search model. 
\end{abstract}


\section{Introduction}

The number of citations to a paper gives an estimate of its quality and relevance
for the scientific community. This fact explains our interest to know how many times
our papers are cited. Besides this absolute measure, we are also interested in the
number of citations relative to other scientist works. For instance, one may look
for the top-cited articles or scientists in a given period. 

Two recent works have reported a statistical analysis of citation data
\cite{laherre98,redner98}. Lah\`erre and Sornette \cite{laherre98} have analyzed the
statistics of the number of citations to a given scientist, within the top-cited
scientists over the period 1981-June 1997. They observed that the distribution of
this magnitude is well fitted by a stretched exponential. On the other hand, Redner
\cite{redner98} has studied the statistics of citations to a given paper using two
different data sets. In this case, for large number of citations the data was fitted
better by a power law decay with exponent 3. Latter Tsallis and Albuquerque
\cite{tsallis00} have claimed that this data can be fitted by a single curve
derived from the non-extensive formalism, with a power law decay for large number
citations.

In both cases they have investigated the in-degree distribution of the citation
network, where the in-degree is the number of citations to a given scientist or a
paper, respectively. However, up to the author knowledge, no report has been done
about the out-degree distribution, where the out-degree is the number of papers
cited within a paper. This magnitude also gives information about the topology of
the network. For instance, since the average in-degree and out-degree are identical,
it is possible to compute the average in-degree from measurements of the out-degree
distribution. The advantage of this approach is that it is easier to collect data
about the out-degree distribution. 

In order to investigate the out-degree distribution of the citation network data has
been collected from different journals in the period 1991-1999. The analysis of the
out-degree distribution reveals that the journals are divided into two classes,
associated with the existence or not of a restriction in the maximum number of pages
per article. Moreover, it is observed that the mean out-degree grows logarithmically
with the number of published papers. Finally, the data is compared with a recursive
search model \cite{vazquez00} obtaining some agreement. It is concluded that the
recursive search is one of the mechanisms that determine the topology of citation
networks. 

\section{Statistical data}
\label{sec:data}

\subsection{Out-degree distribution}
\label{sec:out-degree}

Data of citations within a paper (out-degree $k_{ou}$) of 12 journals was analyzed. 
The data was collected from the Science Citation Index report \cite{ISI}. They were
selected among the journals with more papers published per year in order to have
good statistics.  The out-degree distribution $P_{ou}(k_{ou})$ was computed for the
periods 1991, 1991-92, 1991-93,... 1991-99 for each journal.

\begin{figure}

\centerline{\psfig{file=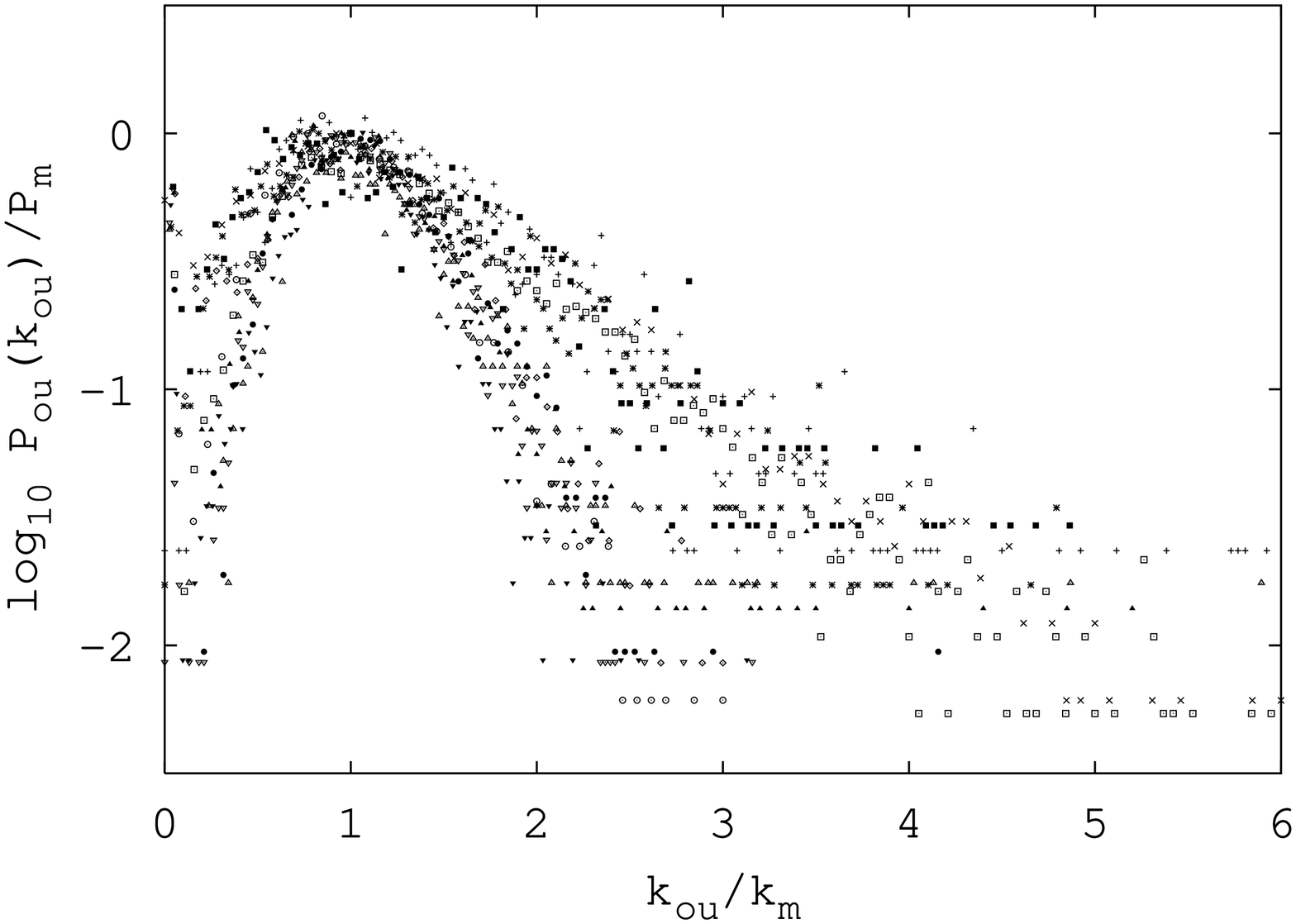,width=5in,angle=-90}}

\centerline{\psfig{file=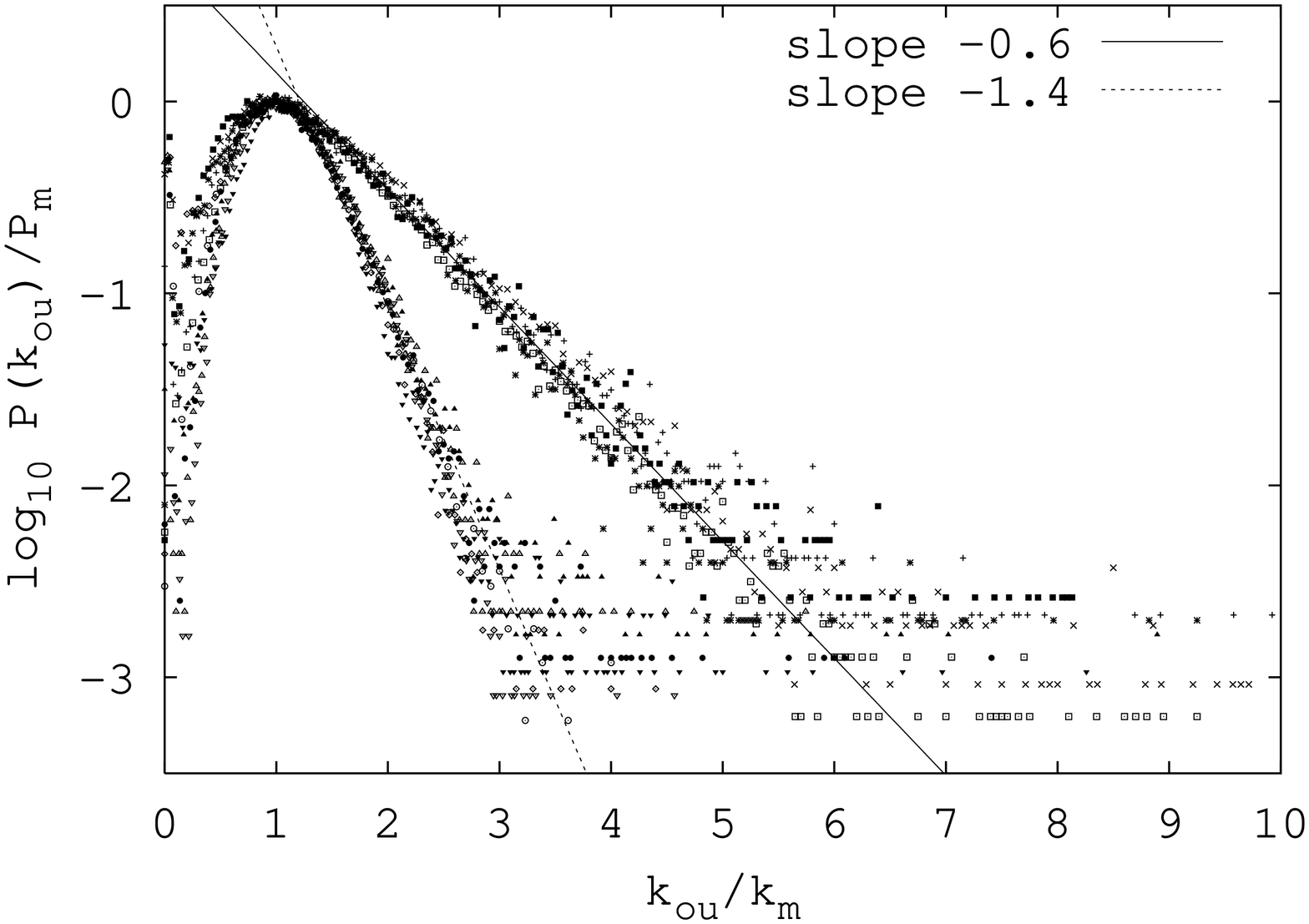,width=5in,angle=-90}}

\caption{Out-degree distribution in the periods 1991(top) and 1991-99 (bottom) for 12
different journals, in a semi-log scale.  The data has been rescaled using the coordinates of
the maximum $(k_m,P_m)$. The symbols correspond to, journals without a restriction in the
maximum number of pages:  Astrophys. J. (plus), J. Appl. Phys. (times), J. Chem.  Phys.
(asterisk), Phys. Rev. B (open square), Phys. Rev. D (filled square); journals with a
restriction in the maximum number of pages: Appl. Phys. Lett. (open circle), Biochem.
Biophys.  Res. Co. (filled circle), Biochem. (USA) (open triangle up), Febs. Lett. (filled
triangle up), J. Biol. Chem. (triangle down), PNAS (filled triangle down), Phys. Rev. Lett.
(diamond). The lines are the best fits to an exponential decay. The slope, in the semi-log
scale, of these exponential decays are indicated in the figure.}

\label{fig:1}
\end{figure}

In all cases the out-degree distribution exhibits a maximum value $P_m$ at an intermediate
out-degree $k_m$. The coordinates of the maximum $(k_m,P_m)$ does not change appreciable from
period to period. The rescaled plots $P_{ou}(k_{ou})/P_m$ vs. $k_{ou}/k_m$ are shown in fig.
\ref{fig:1} for the periods 1991 and 1991-99, while the value of $k_m$ and $P_m$ are given in
table \ref{tab:1}.

For $k<k_m$ there are strong fluctuations of the out-degree distribution from
journal to journal. These fluctuations are not reduced for the 1991-99 period, for which more
statistical data is available. Hence, the shape of the out-degree distribution in this region
is non-universal.

On the contrary, for $k_{ou}>k_m$ all the curves practically overlap into two universal
curves, which divide the journals into two classes. For the 1991 data there is a strong
dispersion in the data and, as a consequence, the universal tendency is not so clear.
Nevertheless, the dispersion is reduced for the 1991-99 data, and the universal behavior
becomes more evident. Hence, there is a generic universal behavior which is obscured by the
statistical dispersion. Moreover, the two universal curves can be fitted by an exponential
decay with decay rates 0.4 and 1.6, respectively.

The analysis of the journals on each class reveals that this subdivision is determined by
the existence or not of a restriction in the number of pages per article. One class
contains the journals with a limitation in the maximum number of pages per article while
the other contains the journals without this restriction. In the first class there is a
faster exponential decay beyond the maximum while the exponential decay is slower for the
other class, {\em i.e.} the restriction in the number of citations within a paper carries
as a consequence a reduction in the width of the out-degree distribution, as one would
expect.

\begin{figure}

\centerline{\psfig{file=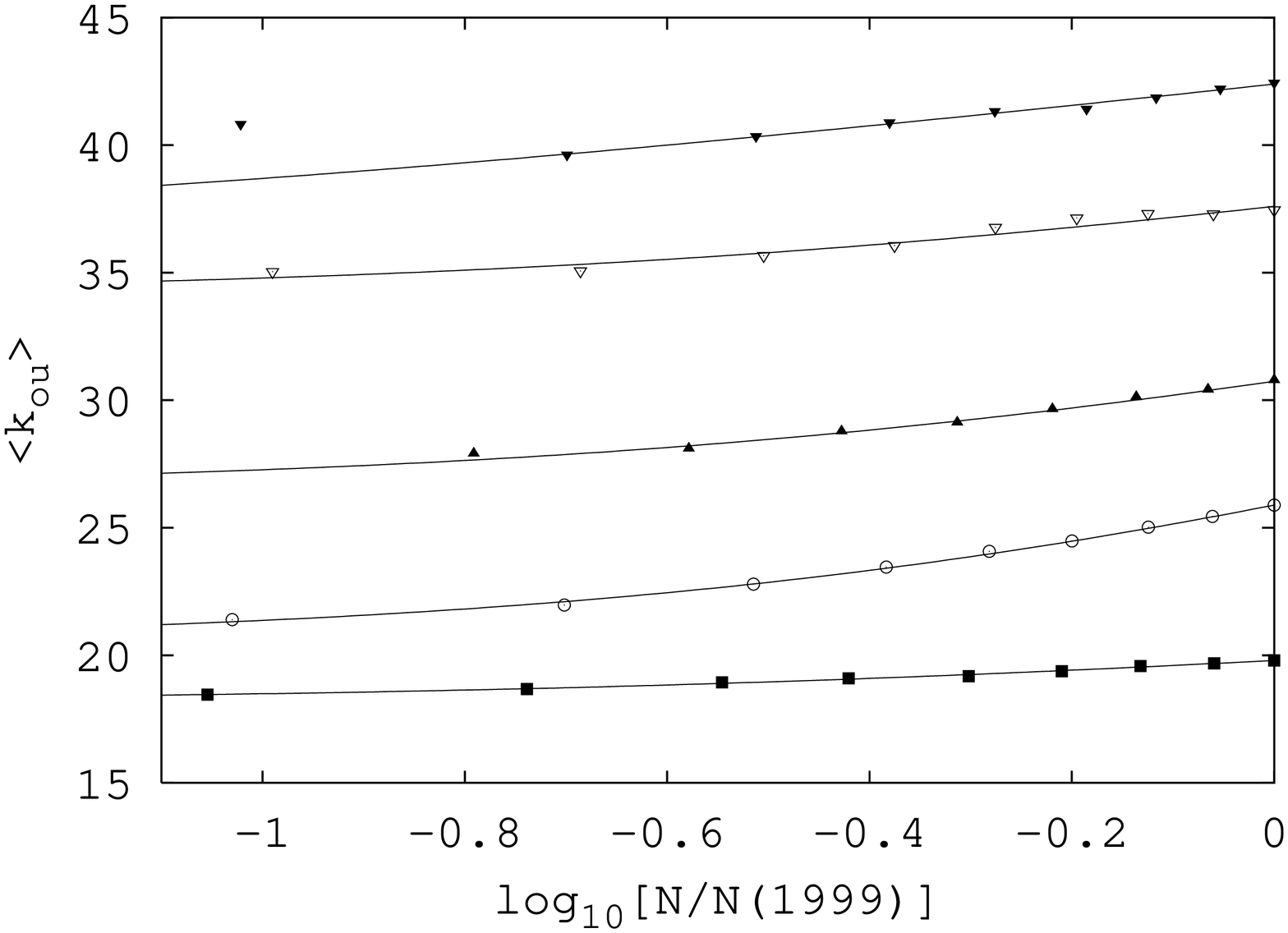,width=5in,angle=-90}}

\centerline{\psfig{file=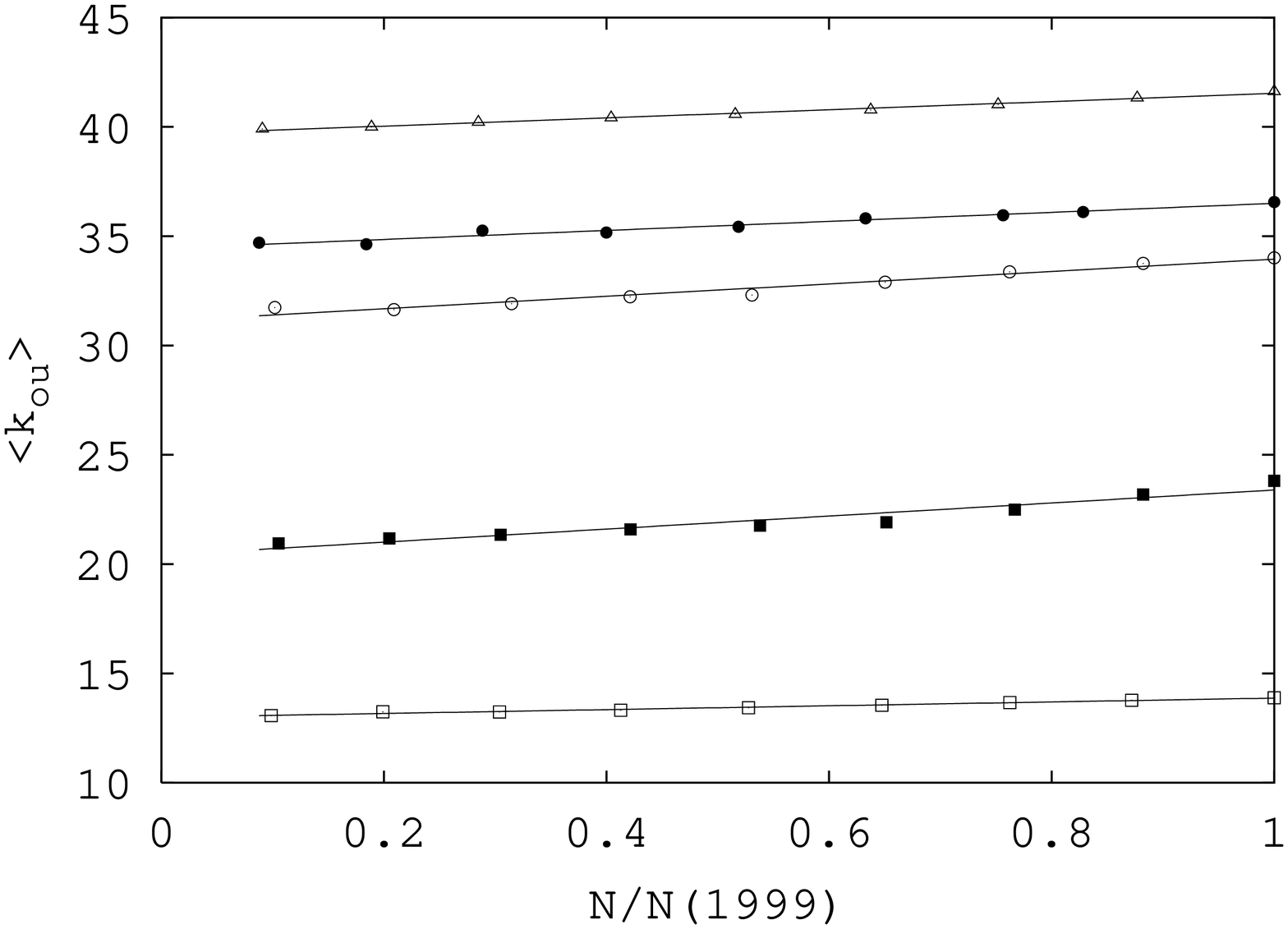,width=5in,angle=-90}}

\caption{Mean out-degree as a function of the number of published papers $N$. Top,
journals which are better fitted by eq. (\ref{eq:2}): Biochem.  (USA) (filled down
triangles), J. Chem. Phys. (open down triangles), Phys. Rev. D (filled up
triangles), Febs. Lett. (open circles), Phys. Rev. Lett. (filled squares). Bottom,
journals which are better fitted by eq. (\ref{eq:3}): J. Biol. Chem. (open up
triangles), Astrophys. J. (filled circles), PNAS (open circles), Biochem.
Biophys. Res. Co. (filled squares), Appl. Phys. Lett. (open squares).}

\label{fig:2}
\end{figure}

The shape of the out-degree distribution around the maximum does not change to much
from period to period and the coordinates of the maximum remains approximately
constant. However, from fig. \ref{fig:1} it can be seen that there are changes in
the tail of the distribution. For 1991 there were just a few papers with
$k_{ou}/k_m$ larger than 1. On the contrary, for the period 1991-99 a lot more were
found. This fact is manifested an the mean out-degree $\langle k_{ou}\rangle$,
which increases slowly from period to period. The plot of $\langle k_{ou}\rangle$
for each period vs. the number of papers $N$ published in that period is shown in
fig. \ref{fig:2}. 

For some journals $\langle k_{ou}\rangle$ is well fitted by a linear dependency with
$N$ while for others a logarithmic growth gives a better fit to the data. These two
different behaviors of $\langle k_{ou}\rangle$ vs. $N$ are not related with the
subdivision into two classes of journals described above. Actually, as it is shown
below, they can be obtained as the limiting cases o a single logarithmic dependency.

If the mean out-degree increases logarithmically with the number of published papers
$n$ (even considering  papers published before 1991) then
\begin{equation}
\frac{d\langle k_{ou}\rangle}{dn}=\frac{b}{n},
\label{eq:1}
\end{equation}
where $b$ is a constant that may change from journal to journal. Now, let $N_0$ be
the number of papers published before 1991 and let $k_0$ be the mean out-degree
considering the $N_0$ published before 1991. Then, integrating eq. (\ref{eq:1}) from
$N_0$ to $N_0+N$ one obtains that
\begin{equation}
\langle k_{ou}\rangle=k_0+b\ln\left(1+\frac{N}{N_0}\right).
\label{eq:2}
\end{equation}
For $N/N_0\ll1$ this expression can be approximated by
\begin{equation}
\langle k_{ou}\rangle\approx k_0+b\frac{N}{N_0},
\label{eq:3}
\end{equation}
while for $N/N_0\gg1$ one obtains
\begin{equation}
\langle k_{ou}\rangle\approx k_0+b\ln\frac{N}{N_0}.
\label{eq:4}
\end{equation}
Hence, the linear and logarithmic growth can be obtained as two asymptotic behaviors
of eq. (\ref{eq:2}) depending on the ratio $N/N_0$. 

The data was fitted by the logarithmic dependency in (\ref{eq:2}). For some journals
(J. Chem. Phys., Phys. Rev. B, Phys. Rev. D, Biochem., Febs. Lett., Phys. Rev. Lett.) the
fitting to that expression is very good obtaining the complete set of parameters $k_0$,
$b$, and $N_0$. However, for the other journals, any large value of $N_0$ yield an
equally good fit to the data, but always with the ratio $b/N_0$ remaining constant. This
suggest that the data for these journals is in the limiting case given by
eq. (\ref{eq:3}), where the linear dependency is observed.

The parameters $N_0$, $b$ and $b/N_0$ resulting from the best fit are reported in table
\ref{tab:1}. The values of $N_0$ are small if one expect that $N_0$ is the total number of
papers published before 1991. For instance for Biochem. (USA) $N_0\approx980$, which is
smaller than the number of papers published per year on this journal (around 1500). Hence,
$N_0$ cannot be interpreted as the total number of papers published before 1991. This fact
suggest that the logarithmic dependency is not manifested in the past history of the
journals.

\subsection{In-degree distribution}
\label{sec:in-degree}

In a directed network the average in-degree and out-degree are identical. 
Nevertheless, we should be careful in extrapolating the results obtained here for
the average out-degree.  The journals have cross-references among them and,
therefore, the in-degree distribution relates different journals. In principle one
should group the journals in clusters in such a way that journals in different
clusters have not cross-references among them. In any case the fact that the growth
has been observed for all the journals analyzed here is a strong evidence that the
overall average out-degree, which is equal to the overall average in-degree,
increases with the number of published papers. Moreover, the growth is well fitted
by the logarithmic dependency in eq. (\ref{eq:2}) or its asymptotic limits in eqs.
(\ref{eq:3}) and (\ref{eq:4}).

We can go beyond averages and make also some conclusions concerning to the in-degree
distribution. If the in-degree distribution follows the power law $P_{in}(k_{in})\sim
k_{in}^{-\gamma}$ and its average grows logarithmically with the network size then the
exponent $\gamma$ should be equal to 2, or at least close to it. Recently Redner
\cite{redner98} have done a statistical analysis of the in-degree distribution of two
different data sets. The first was the citation distribution of papers published in 1981
and cited between 1981-June 1997 (ISI). The second data set was the citation distribution,
as of June 1997, of the papers published in Phys. Rev. D in the period 1974-1994 which were
cited at least once (PRD).

In the ISI data only citations to papers published in 1981 is considered, while
citation to papers published between 1982-June 1997 is not. Hence, This data set is
better to investigate the process of aging in the citation network, in which very
old papers are rarely cited. However, it is not good to investigate the in-degree
distribution because it does not considers the citation to papers published between
1982-June 1997. On the other hand, in the PRD data one is not considering the
citations to papers published in the period 1995-June 1997. However, they are only
three years over a period of 23 years. Hence, with this data set one has a better
approximation to the in-degree distribution. 

\begin{figure}

\centerline{\psfig{file=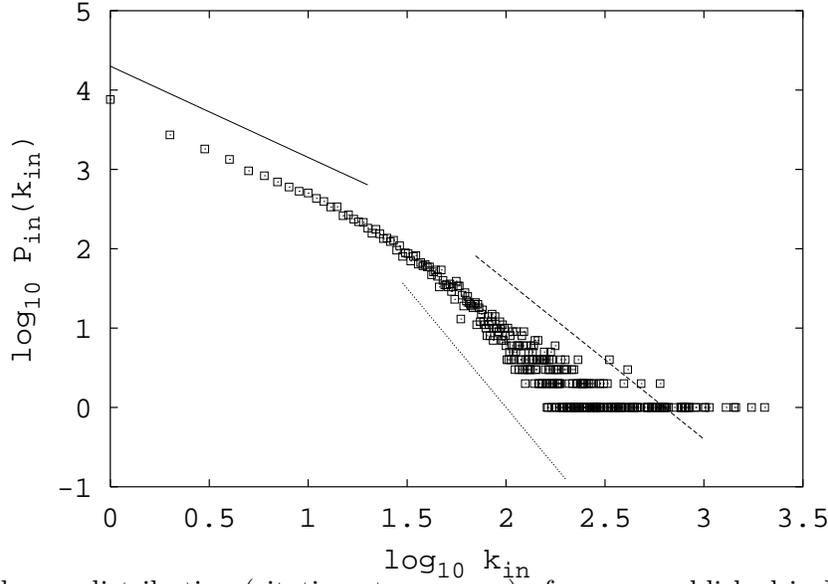,width=4.5in,angle=-90}}

\caption{In-degree distribution (citations to a paper) of papers published in Phys. Rev.
D in the period 1982-June 1997, in a log-log scale. The straight lines are power laws
$P(k_{in})\sim k_{in}^{-\gamma}$ with exponent : $\gamma=1.3$ (continuous line),
$\gamma=1.9$ (dashed line), and $\gamma=3$ (dotted line).}

\label{fig:2a}
\end{figure}

A log-log plot of the in-degree distribution for the PRD data is shown in fig.  
\ref{fig:2a}. The distribution shows two different regimes. For small in-degrees the
distribution can be fitted by a power law with an approximate exponent 1.3. For large
in-degrees the curve can be also fitted by a power law but with a larger exponent. A power
law fit to this part of the curve gives the exponent $\gamma=1.9\pm0.2$. This exponent is
smaller than the value 3 reported by Redner \cite{redner98}. In any case, this second power
law, is manifested in only one decade and, therefore, is difficult to make any final
statement with the present data. However, the exponent $\gamma=1.9\pm0.2$ reported here is
more consistent with the out-degree distribution data analyzed above.

A power law in-degree distribution with power law exponent $\gamma=3$ will give a finite
average when $N\rightarrow\infty$. For finite $N$ the mean in-degree will grow with $N$,
with a tendency to saturate to the stationary value obtained when $N\rightarrow\infty$. On
the other hand, if the exponent is $\gamma=2$ the mean in-degree will always grow with $N$
following a logarithmic dependency. This second picture is more consistent with the
behavior of the mean out-degree as a function $N$, which increases logarithmically and
in some cases linear with $N$.

\section{Recursive search model}
\label{sec:model}

A second step will be to propose a model that explains the features described above,
{\em i.e.} an average out-degree (in-degree) which increases logarithmically with
the network size, an out-degree distribution with a maximum at intermediates
out-degrees and a power law in-degree distribution with decay exponent 2. In the
literature we can find two models which lead to power law in-degree distributions. 
They are the free-scale and the recursive search models, introduced by Barab\'asi
and co-workers \cite{barabasi99} and by the author \cite{vazquez00}, respectively.

The free-scale model and its generalizations \cite{dorogovtsev00,krapivsky00} lead
to power law in-degree distributions with power law exponent $\gamma>2$ and,
therefore, with a finite average in-degree. The exponent $\gamma=2$ can be reached
asymptotically but only after a fine-tune of the parameters involved in the model. 
On the contrary, the recursive search model \cite{vazquez00} is characterized by a
robust exponent $\gamma=2$ over a wide range of its parameters, leading to a
logarithmic growth of the average in-degree as the network size increases. On the
other hand, in the free-scale model the out-degree is fixed for all nodes. Hence, to
describe the out-degree distribution observed in the citation data, one has first to
make some generalization of it, which introduces some randomness in the out-degree.
On the contrary, as it is shown below, the recursive search model leads to
out-degree distributions that are already in good agreement with the citation data.
All these elements suggest that the recursive search model is more appropriate to
describe the citation network. 

The main idea of a recursive search is to be connected to one node (paper) of the
network and any time we get in contact with a new node follows its links, exploring
in this way part of the network. This process mimics in some elementary way our
recursive searches over the citation network. In order to perform an efficient
search one usually follows only the links of those nodes that appear to be of good
quality according to certain criteria. To take into account this effect the
parameter $p$ ($0\leq p\leq1$) is introduced, which is defined as the probability
that a node satisfies our quality requirements. Thus, it is assumed that each time a
node satisfies our quality requirements, which happens with probability $p$, a link
is created to it and the search is continued latter following its links. This search
process has the limitation that in order to follow the links of a node one has to
create first a link to it, a constraint which may be no realistic in some cases. 
However, it already captures the essential features of a recursive search.

The recursive search model is then defined by two rules: adding and walking. In the
adding rule one adds a node to the network. If more than one node is present then
one performs the walking rule defined as follows: if a link is created to a node in
the network then with probability $p$ a link is also created to each of its nearest
neighbors. This rule is performed until no link is created and in that case one goes
back to the adding rule.

The model exhibits a phase transition at a threshold probability $p_c\approx0.4$
\cite{vazquez00}. Below $p_c$ the average in-degree approaches a stationary value as
the network size (number of nodes $N$) increases. On the contrary, above the
threshold the average out-degree (in-degree) increases logarithmically with $N$ and
the in-degree distribution displays a robust power law with exponent
$\gamma\approx2$, independent of $p$. Thus, in the range $p_c\leq p\leq1$ the
in-degree distribution obtained from the model has the same features discussed above
for the citation data. 

For $p=1$ the average out-degree (in-degree) increases as $\langle
k_{ou}\rangle=a+b\ln N$ with $b=1$. For $p_c<p<1$ it still increases logarithmically
with $N$ but with $b<1$. However, the statistical data reported in the previous
section shows that $b$ is in general greater than 1, ranging from 1.2 to 5.2. 

\begin{figure}

\centerline{\psfig{file=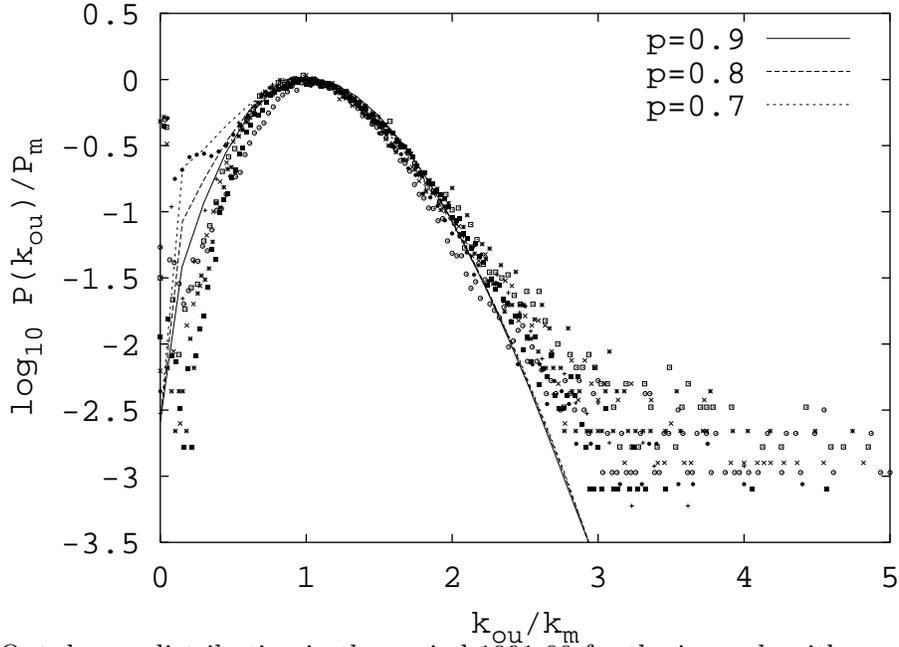,width=5in,angle=-90}}

\caption{Out-degree distribution in the period 1991-99 for the journals with a restriction in
the maximum number of pages: Appl. Phys. Lett. (plus), Biochem.  Biophys. Res. Co. (times),
Biochem.  (USA) (asterisk), Febs. Lett. (open square), J.  Biol. Chem. (filled square), PNAS
(open circle), Phys. Rev. Lett. (filled circle)), in a semi-log scale.  The lines correspond
with the out-degree distribution as obtained form the recursive search model using the values
of $p$ indicated in the figure. As in the real data, these curves have been shifted by the
coordinates of their respective maxima.}

\label{fig:3}
\end{figure}

On the other hand, the comparison of the out-degree distribution obtained from the
model and that reported here for different journals is shown in fig. \ref{fig:3}. As
it can be seen there is some agreement with the data for the class of journals with
a limitation in the number of pages per article. The main features, a maximum at
intermediate out-degrees, universal behavior at the right of the maximum and
non-universal behavior at the left of the maximum, are also observed for the
recursive search model. However, the model underestimate the out-degree distribution
for large $k_{ou}$ and it does not give a good agreement with the journals without
restriction in the number of pages. 

Hence, the recursive search model explains the main features observed for the
in-degree and out-degree distributions of the citation network. However, the model
is still to simple to give a complete description, with a quantitative agreement.
One can make a generalization of the model in which one starts the
search at different nodes. This modification will carry as a consequence a wider
out-degree distribution and probably a better agreement with the journals without a
restriction in the number of pages. On the other hand, the preferential attachment
\cite{barabasi99} is another mechanism which may comes to play. During the search
one may follow preferentially those paper which are more cited, which is a measure
of their quality. This will be manifested in the recursive search model as a
dependence of $p$ on $k_{in}$. Work on that direction is in progress.

\section{Summary and conclusions}
\label{sec:summary}

In summary, empirical data of the out-degree distribution of the network of citations has
been presented. The main features observed were a maximum at intermediate out-degrees,
universal behavior at the right of the maximum, non-universal behavior at the left of the
maximum, and an average out-degree (in-degree) which increases logarithmically with the
number of published papers. These features were in part explained using a recursive search
model introduced in \cite{vazquez00}. It is then concluded that recursive searches are one
of the fundamental mechanisms in the formation of citation networks.

\section*{acknowledgments}

I thanks Y. Moreno Vega and A. Vespignani for helpful comments and suggestions.

\begin{center}
\begin{table}
\begin{tabular}{|l|l|l|l|l|l|}
\hline
Journal & $k_m$ & $P_m$ & $N_0$ & $b$ & $b/N_0$\\
\hline
Astrophys. J.$^{\ref{eq:3}}$ & 26 & 0.026 & - & - & 1e-4\\
J. Appl. Phys.$^{\ref{eq:3}}$ & 14 & 0.044 & - & & 7e-5\\
J. Chem. Phys.$^{\ref{eq:2}}$ & 28 & 0.024 & 7700 & 2.6 & 4e-4\\
Phys. Rev. B$^{\ref{eq:2}}$ & 20 & 0.038 & 33000 & 2.2 & 7e-5\\
Phys. Rev. D$^{\ref{eq:2}}$ & 23 & 0.028 & 5900 & 3.5 & 6e-4\\
Appl. Phys. Lett.$^{\ref{eq:3}}$ & 13 & 0.079 & - & - & 4e-5\\
BBRC$^{\ref{eq:3}}$ & 22 & 0.048 & - & - & 2e-4\\
Biochem. (USA)$^{\ref{eq:2}}$ & 39 & 0.027 & 980 & 1.9 & 2e-3  \\
Febs. Lett.$^{\ref{eq:2}}$ & 24 & 0.043 & 7700 & 5.2 & 7e-4\\
J. Biol. Chem.$^{\ref{eq:3}}$ & 37 & 0.030 & - & - & 5e-5\\
PNAS$^{\ref{eq:3}}$ & 31 & 0.039 & - & - & 1e-4\\
Phys. Rev. Lett.$^{\ref{eq:2}}$ & 20 & (\ref{eq:2}) & 8300 & 1.2 & 1e-4\\
\hline
\end{tabular}

\caption{Coordinates of the maximum ($k_m,P_m$) for each journal and fitting parameters
obtained from the fit of the mean out-degree to either eqs.  (\ref{eq:2}) or (\ref{eq:3}). On
each case the equation used to fit the data is indicated as a supra-script.}

\label{tab:1}
\end{table}
\end{center}

\end{document}